\documentclass[12pt]{article}
\usepackage{a4wide,amssymb,cite}
\pdfoutput=1

\usepackage{a4wide,amssymb,graphicx}
\usepackage{epsfig}
\usepackage[usenames,dvipsnames]{color}
\usepackage{slashed}
\usepackage{amssymb,cite,graphicx}
\usepackage{slashed}
\usepackage{amsmath,bm,bbm}
\usepackage{amsfonts}
\usepackage[titletoc,title]{appendix}
\usepackage[small]{caption}
\usepackage[margin=1in]{geometry}
\usepackage[multiple]{footmisc}
\usepackage{mathtools}
\usepackage{slashed}
\usepackage[nottoc]{tocbibind}
\usepackage{xcolor}

\newcommand{\be}{\begin{equation}}
\newcommand{\ee}{\end{equation}}
\newcommand{\bea}{\begin{eqnarray}}
\newcommand{\eea}{\end{eqnarray}}

\def\circa#1{\,\raise.3ex\hbox{$#1$\kern-.75em\lower1ex\hbox{$\sim$}}\,}

\begin{document}

\begin{titlepage}

\rightline{CERN-TH-2026-219}

\begin{centering}
\vspace{1cm}
{\Large {\bf Inelastic dark matter and baryon flavor symmetry \vspace{0.2cm} \\ in light of LUX-ZEPLIN (LZ) experiment}} \\

\vspace{1.5cm}

\begin{centering}
{\bf  Hyun Min Lee}
\end{centering}
\\
\vspace{.5cm}

{\it Department of Physics, Chung-Ang University, Seoul 06974, Korea.}\vspace{0.2cm} \\
{\it Theoretical Physics Department, CERN, 1211 Geneva, Switzerland.}

\vspace{.5cm}


\end{centering}
\vspace{2cm}

\begin{abstract}
\noindent
We present a novel model for inelastic dark matter in a simple extension of the Standard Model (SM) with a local $U(1)'$ baryon flavor symmetry, such as $U(1)_{B_i-B_j}$ with $i\neq j$. In this framework, we realize a realistic flavor structure for the mixing Yukawa couplings due to higher dimensional operators or  renormalizable couplings between the SM quarks and extra vector-like quarks that are anomaly-free by themselves.  We regard a Dirac singlet fermion as being dark matter and obtaining a small mass splitting for inelastic dark matter, only after the $U(1)'$ symmetry is broken spontaneously.
We show that there is a consistent parameter space for explaining both the LZ nuclear recoil event and the correct relic density for dark matter and the existing collider and indirect detection bounds are satisfied at the same time. In this case, we predict a relatively light $Z'$ boson with $10-20\,{\rm GeV}$ mass and a weak gauge coupling, suggesting a dedicated search at colliders. The future data with higher recoil energies at direct detection experiments will play a decisive role of discriminating between the scenarios of light and heavy dark matter.

\end{abstract}

\vspace{2.5cm}

\begin{flushleft} 
$^\dagger$Email: hminlee@cau.ac.kr
\end{flushleft}

\end{titlepage}

\section{Introduction}

An interesting indication for dark matter (DM) signal has been announced from a nuclear recoil at LUX-ZEPLIN (LZ) experiment \cite{LZ}. There is one event showing up at $E_R=248\pm 23({\rm stat})\pm 23({\rm sys})\,{\rm keV}$, which is in the extended nuclear recoil energy window, as compared to the case for elastic DM-nucleon scattering for which there  exist strong limits for dark matter by LZ \cite{LZ-elastic}, XENON \cite{Xenon} and  Panda-X experiments \cite{PandaX}, etc.

Inelastic dark matter scenarios \cite{inelastic,electron-inelastic} can provide a new avenue for dark matter searches because the mass gap between two distinct dark matter particles is sufficiently large to avoid the strong  bounds on elastic DM-nucleon scattering from direct detection experiments.
Moreover, there is an interesting possibility of detecting dark matter through inelastic DM-nucleon scattering, for which  the observable window is extended to higher recoil energies in the presence of a large DM velocity in the DM halo. 

Higgsino dark matter with $1\,{\rm TeV}$ mass can account for the LZ nuclear recoil event for a small mass gap between two superpartners of Higgs bosons in the supersymmetric extension of the Standard Model (SM), satisfying the relic density condition simultaneously \cite{Higgsino}. There are still issues in interpreting the LZ data: null observation of even higher recoil energy events expected from inelastic dark matter \cite{Safdi}, compatibility with other experiments such as IceCube for neutrino searches from the Sun \cite{Pospelov,Icecube2}. Nonetheless, there are a plethora of immediate responses from the theory community to the LZ excess in terms of models for inelastic dark matter with endothermic \cite{Higgsino,endothermic} or exothermic \cite{exothermic} processes, elastic dark matter with axion-like mediator \cite{alps}, dark matter absorption and other possibilities \cite{absorption,papers}. 

In this article, we propose a new model for inelastic dark matter in the presence of extra local $U(1)'$ symmetries coupled to baryonic currents composed of SM quarks. We consider $U(1)_{B_i-B_j}$ with $i\neq j$  as concrete examples for anomaly-free baryon flavor symmetries and require at least the first generation quarks to carry nonzero $U(1)'$ charges for the DM-nucleon scattering at the LZ experiment.  We introduce  a Dirac singlet fermion for dark matter, which receives a small Majorana mass after the $U(1)'$ is broken spontaneously by the VEV of a singlet scalar field carrying a nonzero $U(1)'$ charge.  The Dirac singlet fermion is then split into two singlet fermions, $\chi_1$ and $\chi_2$, and the former is slightly lighter than the latter.

We look for the consistent parameter space in our model to explain the LZ excess through the inelastic DM-nucleon scattering, namely, $\chi_1 N\to \chi_2 N$. Moreover, we take into account the generation of quark flavor mixings in the model and other constraints on the model, coming from the DM relic density, dijet bounds, indirect detection bounds, etc.

The paper is organized as follows. 
We first describe the model setup for inelastic dark matter and $U(1)$ baryon flavor symmetries, and identify the form of the Yukawa couplings and new $Z'$ interactions of dark matter and SM quarks.  
We recap the general discussion on inelastic dark matter by kinematics and event rate calculations and apply the result to the interpretation of  the LZ data in our model. 
Next we discuss how the DM relic density condition, existing dijet bounds, indirect direction for DM annihilations can constrain our model. Finally, conclusions are drawn.

\section{Model for inelastic dark matter and baryon flavor symmetry}

In this section we introduce the model setup for inelastic dark matter and the communication of dark matter to the SM via the baryonic currents.

We introduce an extra local $U(1)'$ symmetry with the associated gauge boson $Z'$ and it is broken spontaneously by the VEV of a complex singlet scalar field $\phi$, which is charged under the $U(1)'$.
A Dirac singlet fermion $\chi$ is also charged under the $U(1)'$, so it has both the $U(1)'$-invariant Dirac mass and the Majorana mass induced by the $U(1)'$ symmetry breaking. Thus, for a small Majorana mass for $\chi$, we can realize the inelastic dark matter scenario with a small mass splitting.

There are various realizations of the anomaly-free $U(1)'$ symmetry without additional fermions. But, for DM-nucleon scattering in direct detection experiments, we focus on the case where $Z'$ couples to the SM fermions $f$ through a baryonic current, such as $U(1)'=U(1)_{B_i-B_j}$ with $i\neq j$ or $U(1)'=U(1)_{B_i+B_j-2B_k}$ with $i,j,k$ being all different \cite{quarkflavor}.  We note that a nonzero gauge kinetic mixing with the hypercharge gauge boson, $Z'_{\mu\nu} B^{\mu\nu}$, can be introduced in this case, but it can be negligible when the $U(1)'$ symmetry is embedded in the $SU(3)_q$ quark flavor symmetry \cite{quarkflavor}. Thus, we take the gauge kinetic mixing to zero in our work.

The Lagrangian of the model is given by the following,
\bea
{\cal L}&=&i{\bar\chi}\gamma^\mu D_\mu\chi -m_\chi {\bar\chi}\chi - (y\phi \overline{\chi^c}\chi +{\rm h.c.}) \nonumber \\
&&+ |D_\mu\phi|^2- V(\phi)  -\frac{1}{4} Z'_{\mu\nu} Z^{\prime\mu\nu}  +\sum_f i{\bar f}\gamma^\mu D_\mu f,
\eea
with
\bea
D_\mu\chi&=& (\partial_\mu+i q_\chi g_{Z'}Z'_\mu)\chi,  \\
D_\mu \phi &=& (\partial_\mu-2i q_\chi g_{Z'} Z'_\mu)\phi, \\
D_\mu f &=& \Big(\partial_\mu +i (B_i-B_j) g_{Z'} Z'_\mu\Big)f.
\eea
Here, the potential for the singlet scalar is given by $V(\phi)=-m^2_\phi |\phi|^2+\lambda_\phi |\phi|^4$ and we can also add a Higgs-portal term, $\lambda_{\phi H}|\phi|^2 |H|^2$, in the scalar potential. Here, we took the $U(1)'$ charge for $\phi$ as $q_\phi=-2q_\chi$ for the Majorana mass term.

\subsection{The quark Yukawa couplings}

First, we find that the $U(1)'$ symmetry allows for the diagonal Yukawa couplings for quarks, with $k\neq i, j$, as follows,
\bea
-{\cal L}_{Y,0}&=& y_{d,ii}{\bar q}_i H d_{iR} + y_{d,jj}{\bar q}_j H d_{jR} +y_{d,kk}{\bar q}_k H d_{kR} \nonumber \\
&&+y_{u,ii}{\bar q}_i {\tilde H} u_{iR} + y_{u,jj}{\bar q}_j {\tilde H} u_{jR} +y_{u,kk}{\bar q}_k {\tilde H} u_{kR} +{\rm h.c.}
\eea
Moreover, the quark flavor mixing between $f_i$ and $f_j$ can be generated from the higher dimensional operators containing $\phi$, 
\bea
-{\cal L}_{Y, 1}&=&a_{d,ij}\frac{\phi^2}{M^2}\, {\bar q}_i H d_{jR}+a_{u,ij}\frac{\phi^2}{M^2}\, {\bar q}_i {\tilde H} u_{jR}+b_{d,ij}\frac{(\phi^*)^2}{M^2}\, {\bar q}_j H d_{iR}+b_{u,ij}\frac{(\phi^*)^2}{M^2}\, {\bar q}_j {\tilde H} u_{iR}  +{\rm h.c.} \label{mix1}
\eea
for $q_\phi=B_i$ or $q_\chi=-\frac{1}{2}B_i$.  For $i=3$ and $j=1$, a small mixing angle between the first and second generation quarks can be obtained from the dimension-6 operators for $\langle\phi\rangle/M\ll 1$ in this framework.
We also list the rest of the quark flavor mixings from the other higher dimensional operators  as
\bea
-{\cal L}_{Y,2}&=&a_{d,ki}\frac{\phi^{*}}{M}\, {\bar q}_k H d_{iR}+a_{u,ki}\frac{\phi^{ *}}{M}\, {\bar q}_k {\tilde H} u_{iR}  +b_{d,ik}\frac{\phi^{ }}{M}\, {\bar q}_i H d_{kR}+b_{u,ik}\frac{\phi^{}}{M}\, {\bar q}_i {\tilde H} u_{kR} \nonumber \\
&+&a_{d,kj}\frac{\phi^{ }}{M}\, {\bar q}_k H d_{jR}+a_{u,kj}\frac{\phi^{ }}{M}\, {\bar q}_k {\tilde H} u_{jR} +b_{d,jk}\frac{\phi^{ *}}{M}\, {\bar q}_j H d_{kR}+b_{u,jk}\frac{\phi^{ *}}{M}\, {\bar q}_j {\tilde H} u_{kR} +{\rm h.c.} \label{mix2}
\eea
 For $i=3$ and $j=1$, a small Cabibbo angle can be obtained by the dominant dimension-5 operators as above  for $\langle\phi\rangle/M\ll 1$. Thus, for the connection to the quark flavor mixings, we focus on the $U(1)_{B_3-B_1}$ symmetry in the later discussion. After electroweak symmetry breaking, we obtain the quark mass matrices, but there are flavor-changing neutral currents (FCNCs) due to $Z'$ in the basis of mass eigenstates for quarks, because of the flavor-dependent couplings for $Z'$. Thus, there are strong constraints on FCNCs due to $Z'$, but we don't pursue this important issue in the current paper.

The non-renormalizable operators in eqs.~(\ref{mix1}) and (\ref{mix2}) can be attributed to vector-like quarks that are $U(1)'$ anomaly-free by themselves and have the renormalizable Yukawa couplings to quarks. Suppose that the  $U(1)'$-invariant renormalizable vector-like quarks, $Q_L, Q_R$, carrying the $U(1)'$ charge, $q_Q=-B_i$, are given by
\bea
-{\cal L}_Q=M_Q {\bar Q} Q + (\lambda'_{d,i} {\bar Q}_L H d_{iR}+\lambda'_{u,i} {\bar Q}_L {\tilde H} u_{iR}  + \lambda_{q,k}\phi\, {\bar q}_k Q_R+{\rm h.c.}).
\eea
Then, integrating out the vector-like quark, we obtain the effective leading mixing Yukawa couplings, as follows,
\bea
{\cal L}_{Y,{\rm eff}}=\frac{\lambda_{q,k}\lambda'_{d,i}}{M_Q}\, \phi\, {\bar q}_k H d_{iR} +\frac{\lambda_{q,k}\lambda'_{u,i}}{M_Q}\, \phi\, {\bar q}_k {\tilde H} u_{iR}+{\rm h.c.},
\eea
identifying 
\bea
\frac{a_{d,ki}}{M}= -\frac{\lambda_{q,k}\lambda'_{d,i}}{M_Q}, \quad \frac{a_{u,ki}}{M}= -\frac{\lambda_{q,k}\lambda'_{u,i}}{M_Q}.
\eea

We remark that it is more flexible to choose types of $U(1)'$ in the presence of an extra singlet scalar $\phi'$ carrying the $U(1)'$ charge, $q_{\phi'}=2B_i$. In this case, the quadratic couplings for $\phi$ in eq.~(\ref{mix1}) can be replaced by the linear couplings for $\phi'$ such that the $U(1)_{B_2-B_1}$  symmetry can be also consistent with the realistic flavor structure.
Therefore, for direct detection at LZ, we consider either $U(1)_{B_3-B_1}$ or $U(1)_{B_2-B_1}$ in the following discussion.

\subsection{Extra gauge interactions for inelastic dark matter}

After the singlet scalar $\phi$ gets a VEV as $\langle\phi\rangle=\frac{1}{\sqrt{2}}\,v_\phi$, the squared mass for the $Z'$ gauge boson is given by $m^2_{Z'}=q^2_\phi g^2_{Z'} v^2_\phi= B^2_i g^2_{Z'} v^2_\phi$. 
Then, in the basis of $\psi\equiv (\chi, \chi^c)^T$, the mass matrix for the dark fermions is given by
\bea
{\cal L}_{\rm mass}=-\frac{1}{2} {\bar\psi}_R {\cal M}_\psi \psi_L +{\rm h.c.},
\eea
with
\bea
{\cal M}_\psi =\left(\begin{array}{cc} m_\chi & \sqrt{2}y v_\phi \\  \sqrt{2}y v_\phi & m_\chi \end{array} \right).
\eea
Then, we can diagonalize the above Lagrangian  by a rotation of the dark fermions, 
\bea
\left(\begin{array}{c} \chi_1 \\ \chi_2 \end{array} \right) =\frac{1}{\sqrt{2}}\left( \begin{array}{cc}  1 & -1  \\  1 & 1 \end{array}\right) \left(\begin{array}{c} \chi \\ \chi^c \end{array} \right),
\eea
so we get the mass eigenvalues as
\bea
m_{\chi_{1,2}} = m_\chi \mp \sqrt{2} y v_\phi. \label{DMmasses}
\eea
As a result, the mass splitting between two dark fermions is given by $\Delta m = m_{\chi_2}-m_{\chi_1}=2\sqrt{2} y v_\phi$.
Then, for $y v_\phi\ll m_{\chi}$, the mass splitting is small enough for explaining the LZ nuclear energy by the inelastic DM scattering.

Furthermore, rescaling the dark fermions by $\chi_{1,2}\to \sqrt{2}\chi_{1,2}$, we get the $Z'$ gauge interactions for the dark fermion in the basis of mass eigenstates, as follows,
\bea
{\cal L}_{\chi, Z'} = - q_\chi g_{Z'}  \Big({\bar\chi}_1\gamma^\mu \chi_2+{\bar\chi}_2\gamma^\mu \chi_1 \Big) Z'_\mu.
\eea
On the other hand, the $Z'$ gauge interactions for the SM quarks are given by
\bea
{\cal L}_{f, Z'} =-g_{Z'} \sum_q Q_q {\bar q}\gamma^\mu q Z'_\mu,
\eea
with
\bea
Q_t&=&Q_b=-Q_u=-Q_d=\frac{1}{3}, \qquad U(1)'=U(1)_{B_3-B_1}, \\
Q_c&=&Q_s=-Q_u=-Q_d=\frac{1}{3}, \qquad U(1)'=U(1)_{B_2-B_1},
\eea
and others being zero in each case. As a result, we obtain the effective interactions between inelastic dark matter and nucleons in common for both $U(1)_{B_3-B_1}$ and $U(1)_{B_2-B_1}$, as follows,
\bea
{\cal L}_{N,{\rm eff}}= \frac{q_\chi g^2_{Z'}}{m^2_{Z'}}\, ({\bar\chi}_2 \gamma^\mu \chi_1) \sum_{N=p,n} {\bar N}\gamma_\mu N. \label{DM-nucleon}
\eea
Here, we note that loop corrections with heavy quarks do not give rise to the effective interactions of dark matter to gluons for vectorial couplings to $Z'$ due to the Landau-Yang theorem \cite{electron-inelastic,DD-vector}, so we can ignore the contributions of heavy quarks for direct detection.

\section{LZ nuclear recoil from inelastic dark matter}

We revisit the kinematics for DM-nucleon inelastic scattering and the calculation of the event rates for the nuclear recoil at direct detection experiments. Then, we identify some benchmark points for explaining the LZ excess by the inelastic scattering with baryonic currents in our model.

\subsection{Kinematics for inelastic DM scattering}

We consider the inelastic scattering between dark matter and nucleus, $\chi_1 A\to \chi_2 A$.
From the energy conservation,
\bea
E_R=\frac{1}{2}m_{\chi_1} (v^2-v^{\prime 2}) -\Delta m,
\eea
where $v,v'$ are the DM velocity before and after scattering in the lab frame for detectors,
the inelastic scattering is kinematically blocked for a small DM velocity, but a large nuclear recoil energy is possible for a large initial kinetic energy for $\chi_1$.

Using the energy-momentum conservation and $\Delta m=m_{\chi_2}-m_{\chi_1}>0$ and $\Delta m\ll m_{\chi_1}\equiv m_\chi$, we obtain   
\bea
2E_0 - \Delta m -\frac{m_A}{\mu_A}\, E_R =2\sqrt{E_0 (E_0-\Delta m-E_R)}\, \cos\theta_{\rm lab} \label{kinematics}
\eea
where $E_0=\frac{1}{2} m_{\chi} v^2$ is the initial kinetic energy for $\chi_1$, $E_R$ is the nuclear recoil energy and $\theta_{\rm lab}$ is the scattering angle in the lab frame. Here, $m_A$ is the nucleus mass and $\mu_A=m_{\chi} m_A/(m_{\chi}+m_A)$ is the reduced mass for the DM-nucleus system. For instance, for $^{131}{\rm Xe}$, $m_A\simeq 131 m_p$ with $m_p=0.938\,{\rm GeV}$ being the proton mass. Solving eq.~(\ref{kinematics}) for $E_R$, we obtain
\bea
E_R&=&\frac{1}{m^2_A} \Big( (2E_0-\Delta m)\mu_A m_A -2 E_0 \mu^2_A \cos^2_{\rm lab}\Big)\pm \frac{2}{m^2_A} \Big(E_0 \mu^2_A \cos^2\theta_{\rm lab}\Big)^{1/2}\times  \nonumber \\
&&\times\Big[E_0\mu^2_A \cos^2\theta_{\rm lab}+E_0 m_A(m_A-2\mu_A)+m_A(\mu_A-m_A)\Delta m \Big]^{1/2}.
\eea
Then, setting $\cos^2\theta_{\rm lab}=1$, from eq.~(\ref{kinematics}), we can find a possible range of the nuclear recoil energy for the DM velocities below the escape velocity for a given $\Delta m$.

For $\cos^2\theta_{\rm lab}=1$, we also note that the minimal kinetic energy or velocity for dark matter for a given $E_R$ is given by
\bea
v_{\rm min}=\frac{\Delta m+ \frac{m_A E_R}{\mu_A}}{\sqrt{2m_A E_R}}.
\eea
Therefore, we need to take the DM velocity in the DM halo to be greater than $v_{\rm min}$ for a given $E_R$.
We note that the minimum value of $v_{\rm min}$ is given by $v^{\rm min}_{\rm min}=\sqrt{\frac{2\Delta m}{\mu_A}}$, so the larger $\mu_A$ (namely, the heavier the detector nucleus such as tungsten, lead, and uranium), the smaller value needed for $v^{\rm min}_{\rm min}$. For a given detector nucleus, we find that the larger $\Delta m$, the larger value of $v^{\rm min}_{\rm min}$ is required. Thus, we need to take the value of $\Delta m$ for $v^{\rm min}_{\rm min}$ not to exceed the escape velocity in our galaxy.

\begin{figure}[!t]
\begin{center}
 \includegraphics[width=0.46\textwidth,clip]{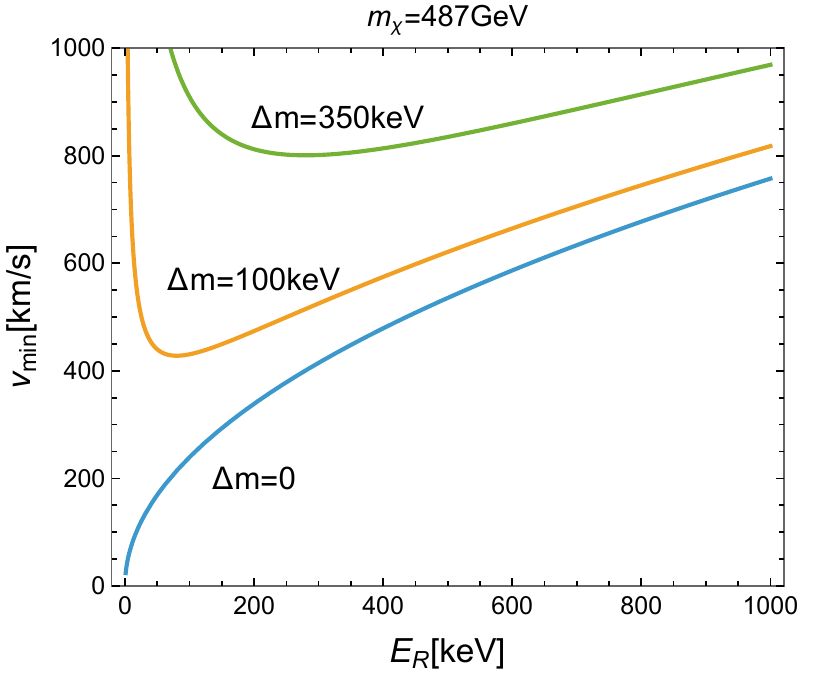} \,\,\,\,
  \includegraphics[width=0.46\textwidth,clip]{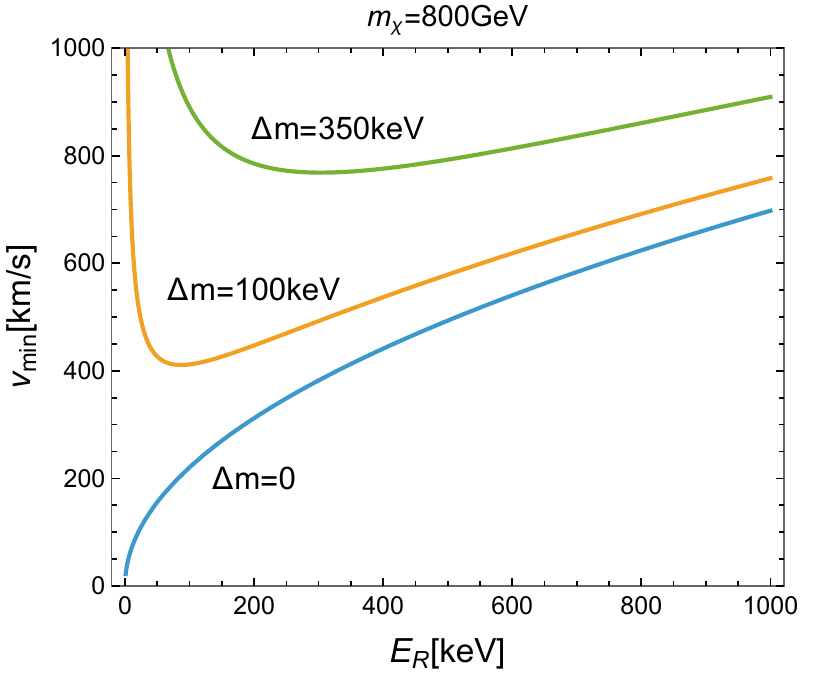} 
 \end{center}
\caption{The minimum DM velocity  $v_{\rm min}$ as a function of the recoil energy $E_R$, for $\Delta m=0, 100\,{\rm keV}, 350\,{\rm keV}$, from bottom to top. We took $m_\chi=487\,{\rm GeV}, 800\,{\rm GeV}$ on left and right, respectively. }
\label{ER}
\end{figure}

In Fig.~\ref{ER}, we show the minimum DM velocity required for the scattering event at the recoil energy $E_R$ as a function of $E_R$. We took $m_\chi=487\,{\rm GeV}, 800\,{\rm GeV}$ on left and right, respectively, and the DM mass splitting varies as $\Delta m=0, {\rm 100\,{\rm keV}}, 350\,{\rm keV}$ from bottom to top in each plot. 

\subsection{Event rate for inelastic DM scattering}

Now we consider the differential event rate for direct detection due to inelastic scattering, which takes the standard expression as for elastic scattering, except the kinematic constraint, as follows,
\bea
\frac{dR}{dE_R}=N_T\,\frac{\rho_{\chi}}{m_\chi}\int_{v_{\rm min}} \frac{d^3 v}{v^{2}}\, v\,  f_1\, \frac{d\sigma}{dE_R},
\eea
with
\bea
\frac{d\sigma}{dE_R}= \frac{m_A}{2\mu^2_N v^2}\, \sigma_N\, \frac{(f_pZ + f_n(A-Z))^2}{f^2_n}\, F^2(E_R) P^2(v). \label{diffcross}
\eea
Here, $Z, A-Z$ are the numbers of protons and neutrons inside the nucleus, respectively, $\mu_N=m_N m_\chi/(m_N+m_\chi)$ is the reduced mass for the DM-nucleon system, $\sigma_N$ is the DM-nucleon scattering cross section, $f_{p,n}$ are the DM-nucleon couplings, $F^2(E_R)$ is the Helm form factor \cite{inelastic,review}, and $P^2(v)$ is the phase space factor, given by
\bea
P^2(v)=\sqrt{1-\frac{2\Delta m}{\mu_A v^2}}, \qquad \Delta m \ll m_\chi.
\eea 
We note that the phase factor is well-defined for $v\geq v_{\rm min}$, because
\bea
vP^2(v)\geq v_{\rm min}P^2(v_{\rm min})=\frac{\big| \Delta m -\frac{m_A E_R}{\mu_A}\big|}{\sqrt{2m_A E_R}}.
\eea
We also note that $N_T$ is the total number of atoms in the detector. For instance, for the LZ nuclear recoil \cite{LZ}, $N_T=2.84$ ton-yr and $\rho_{\chi}$ is the local DM density for $\chi_1$ at the Sun, which is given by $\rho_\chi=0.3\,{\rm GeV/cm^3}$.

Furthermore, taking into account the Earth's motion with respect to the galactic frame, we obtain $f_1(v)$ in the Maxwell-Boltzmann distribution form, 
\bea
f_1= f_1(|{\vec v}+{\vec v}_e|)= \frac{1}{N_m}\, v^2\,  e^{-({\vec v}+{\vec v}_e)^2/v^2_0},
\eea
with the normalization factor,
\bea
N_m=v^3_0 \pi^{3/2}\bigg[{\rm erf}\Big(\frac{v_{\rm esc}}{v_0}\Big)-\frac{2v_{\rm esc}}{\sqrt{\pi} v_0}\, e^{-\frac{v^2_{\rm ecp}}{v^2_0}} \bigg].
\eea
Here, $v_0$ is the Sun's speed, given by $v_0=220\,{\rm km}/s$, $v_{\rm esp}$ is the escape velocity in our galaxy, given by $v_{\rm esc}=544\,{\rm km/s}$, $v_e$ is the Earth's speed taken to $v_e=232\,{\rm km/s}$ in spring/autumn, and ${\rm erf}(x)$ is the error function defined by
\bea
{\rm erf}(x) =\frac{2}{\sqrt{\pi}} \int^x_0 dt^2\, e^{-t^2}.
\eea
Taking $v_{\rm max}=v_{\rm esc}+v_e$ to infinity and ignoring the velocity dependences in the DM-nucleon scattering cross section as well as the phase space factor, we obtain the approximate result for the relevant velocity average \cite{inelastic} as
\bea
\int_{v_{\rm min}} \frac{d^3 v}{v^2}\, \frac{f_1}{v} = \frac{1}{2v_e}\bigg[ {\rm erf}\Big(\frac{v_{\rm min}+v_e}{v_0}\Big)-{\rm erf}\Big(\frac{v_{\rm min}-v_e}{v_0}\Big)\bigg].
\eea

\begin{figure}[!t]
\begin{center}
 \includegraphics[width=0.46\textwidth,clip]{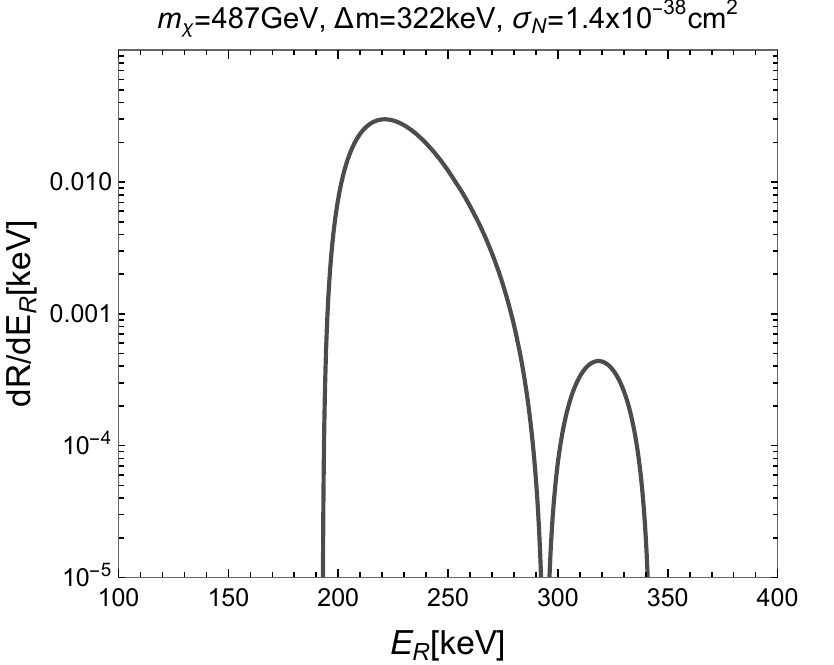}\,\,\,\,
  \includegraphics[width=0.46\textwidth,clip]{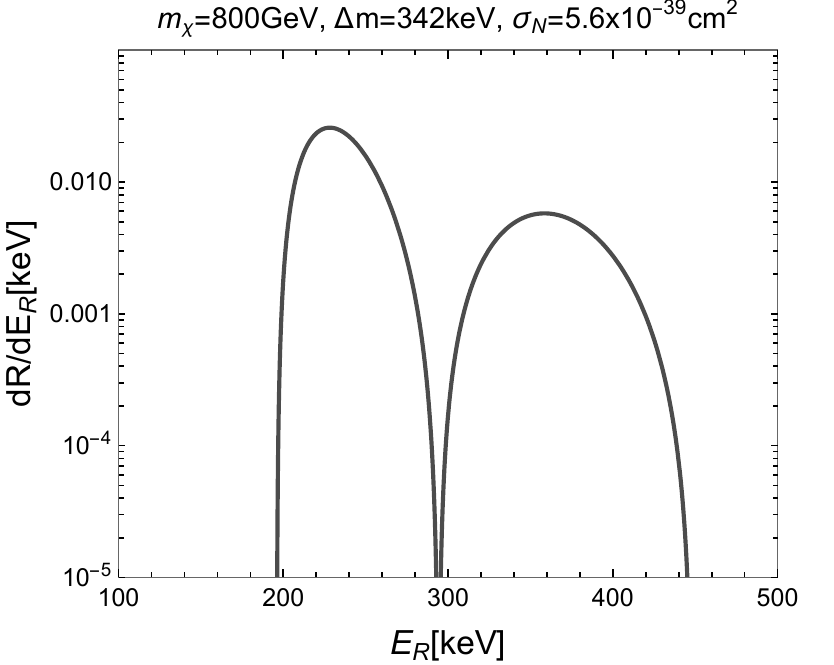}
 \end{center}
\caption{Differential event rates for $N_T=2.84$ ton-yr at LZ experiment, for DM mass, $m_\chi=487\,{\rm GeV}$ on left and $m_\chi=800\,{\rm GeV}$ on right.}
\label{rate}
\end{figure}

As a result, we can compute the total number of events  for scattering as follows,
\bea
R_D=\int_{E_{\rm th}}^\infty \frac{dR}{dE_R}\, dE_R,
\eea
with $E_{\rm th}$ being the detector threshold. In order to compare with the experimental data, we can take the convoluted differential event rate by
\bea
\frac{dR_D}{dE_R} =\frac{R_D}{\sqrt{2\pi}\,\sigma} \, e^{-(E_R-E_c)^2/(2\sigma^2)}.
\eea
For instance, $E_c=248\,{\rm keV}$ and $\sigma=23\,{\rm keV}$ for the LZ event \cite{LZ}. 

Using the LZ nuclear recoil energy, we obtain the total number of events by
\bea
R_D=\int^{E_+}_{E_-} \frac{dR}{dE_R}\, dE_R,
\eea
with $E_\pm= (248\pm 32.5)\,{\rm keV}$ \cite{LZ}. 
In Fig.~\ref{rate}, we depict the differential event rates as a function of $E_R$ for varying the DM mass, the mass splitting as well as the DM-nucleon scattering cross section appropriately. In our case, $f_p=f_n$, and we took $\Delta m=322\,{\rm keV}$ and $\sigma_N=1.4\times 10^{-38}\,{\rm cm}^2$  for $m_\chi=487\,{\rm GeV}$ on left and $\Delta m=342\,{\rm keV}$ and $\sigma_N=5.6\times 10^{-39}\,{\rm cm}^2$  for $m_\chi=800\,{\rm GeV}$ on right. For both of the examples, we set $R_D=1$ for the LZ event. We find that there are sizable events for heavy dark matter at the recoil energies higher than the LZ event. For instance, in the benchmark model with $m_\chi=800(487)\,{\rm GeV}$, the total number of events with recoil energies higher than $E_R=E_+$ is  $R_D(E_R>E_+)=0.4(0.01)$. Thus, we should keep tuned to the potential events in the higher energy bins in the future.

We note that the precise values for the input parameters depend on $v_0$ and $v_{\rm esc}$. 
The larger $v_0$ or $v_{\rm esc}$ \cite{Safdi}, we would need a smaller DM-nucleon cross section for the same DM mass and the DM mass splitting.

\subsection{LZ nuclear recoil and baryon flavor symmetry}

In our model, from the effective interactions between inelastic dark matter and nucleons in eq.~(\ref{DM-nucleon}), we obtain the DM-nucleon scattering cross section entering in eq.~(\ref{diffcross}) by
\bea
\sigma_N=\frac{2\mu^2_N Q^2_q q^2_\chi g^4_{Z'}}{\pi m^4_{Z'}},
\eea
and the effective couplings for nucleons are $f_p=f_n$. Here, we note that $q_\chi=-\frac{1}{2}B_1=-\frac{1}{6}$ and we took $q^2\simeq 2\mu_A v^2\ll m^2_{Z'}$.
As illustration, taking $m_\chi=800\,{\rm GeV}$ and $\sigma_N=5.6\times 10^{-39}\,{\rm cm}^2$ for explaining the LZ event, we need $m_{Z'}=21(10.5)\,{\rm GeV}$ for $g_{Z'}=0.2(0.1)$. For $m_\chi=487\,{\rm GeV}$ and $\sigma_N=1.4\times 10^{-38}\,{\rm cm}^2$, we need $m_{Z'}=16.6(8.3)\,{\rm GeV}$ for $g_{Z'}=0.2(0.1)$.

\section{Other constraints}

We present the condition for the correct relic density and its compatibility with the LZ interpretation.
We also discuss the existing bounds from collider searches, in particular, dijet searches, and comment on other constraints such as flavor violating decays or indirect detection bounds.

\subsection{Relic density}

In our $U(1)'$ models,  the relic densities for dark matter $\chi_1$ and $\chi_2$ are determined by the DM annihilation processes at tree level, $\chi_1{\bar\chi}_2\to q{\bar q}$, $\chi_{1,2}{\bar\chi}_{1,2}\to Z'Z'$, $\chi_1{\bar \chi}_1\to \chi_2{\bar \chi}_2$, $\chi_1\chi_1\to \chi_2\chi_2$, etc. The third and fourth processes are forbidden channels, which are kinematically closed at a zero temperature, but they open up during the freeze-out as far as $T\gtrsim \Delta m$.
On the other hand, the cross sections for $\chi_1$ and $\chi_2$ annihilating into $W^+W^-$ or  $ZZ$ are suppressed due to the loop factors. 

In the limit for non-relativistic dark matter, we list the annihilation cross sections for dark matter after thermal average during the freeze-out. First, the annihilation cross sections for $\chi_1{\bar\chi}_2\to q{\bar q}$, $\chi_1{\bar\chi}_1\to Z'Z'$ are given by
\bea
\langle\sigma v\rangle_{\chi_1{\bar\chi}_2\to q{\bar q}} &=&\frac{N_c q^2_\chi Q^2_q g^2_{Z'}(m^2_q+2m^2_\chi)}{2\pi (4m^2_\chi-m^2_{Z'})^2}\sqrt{1-\frac{m^2_q}{m^2_\chi}},\\
\langle\sigma v\rangle_{\chi_1{\bar\chi}_1\to Z'Z'} &=&\frac{q^2_\chi g^4_{Z'} m^2_\chi}{4\pi (m^2_{Z'}-2m^2_\chi)^2}\bigg(1-\frac{m^2_{Z'}}{m^2_\chi}\bigg)^{3/2}\simeq  \langle\sigma v\rangle_{\chi_2{\bar\chi}_2\to Z'Z'}.
\eea
On the other hand, the annihilation cross sections for $\chi_1{\bar \chi}_1\to \chi_2{\bar \chi}_2$, $\chi_1\chi_1\to \chi_2\chi_2$ are given by
\bea
\langle\sigma v\rangle_{\chi_1{\bar\chi}_1\to \chi_2{\bar\chi}_2} &=&\bigg(\frac{n_{\chi_2}^{\rm eq}}{n^{\rm eq}_{\chi_1}}\bigg)^2\langle\sigma v\rangle_{\chi_2{\bar\chi}_2\to \chi_1{\bar\chi}_1}, \label{selfann1} \\
\langle\sigma v\rangle_{\chi_1{\chi}_1\to \chi_2{\chi}_2} &=&\frac{1}{2}\langle\sigma v\rangle_{\chi_1{\bar\chi}_1\to \chi_2{\bar\chi}_2}.\label{selfann2} 
\eea
Here, $n_{\chi_{1,2}}^{\rm eq}$ are the equilibrium number densities for $\chi_1$ and $\chi_2$, respectively.
The above cross sections take the following form  for $ \Delta m\lesssim T\lesssim m_\chi$, during freeze-out,
\bea
\langle\sigma v\rangle_{\chi_2{\bar\chi}_2\to \chi_1{\bar\chi}_1}=\frac{4 q^4_\chi g^4_{Z'} m^2_\chi}{\pi^{3/2} m^4_{Z'}}\, \sqrt{\frac{T}{m_\chi}},
\eea
while they become, for $T\lesssim \Delta m$,
\bea
\langle\sigma v\rangle_{\chi_2{\bar\chi}_2\to \chi_1{\bar\chi}_1}=\frac{q^4_\chi g^4_{Z'} m^2_\chi}{2\pi m^4_{Z'}}\sqrt{\frac{2\Delta m}{m_\chi}}. \label{selfann3}
\eea
We note that we used the detailed balance condition in deriving eqs.~(\ref{selfann1}) and (\ref{selfann2}).
Then, it is important to notice that the forbidden channel is not Boltzmann-suppressed for $T\gtrsim \Delta m$ and it is enhanced for the small values of $m_{Z'}/g_{Z'}$ as favored by the LZ excess, making $\chi_1, \chi_2$ in chemical equilibrium during freeze-out. But, $\chi_1{\bar \chi}_1\to \chi_2{\bar \chi}_2$, $\chi_1\chi_1\to \chi_2\chi_2$, etc, do not influence the relic densities of $\chi_1$ and $\chi_2$ with a small mass splitting. As a result, we find that a relatively heavy dark matter is required for the LZ event, so the relic density cannot be explained for the parameter space favored by the LZ event, requiring an extension of the model for the correct relic density \footnote{We also note that the Yukawa coupling for DM fermion is negligible for the relic density calculation, because a small Yukawa coupling $y$ is needed for a small mass splitting, as discussed below eq.~(\ref{DMmasses}).}.

We now consider a simple extension of the model with a singlet real scalar $\phi'$ with a new Yukawa coupling to dark matter, 
\bea
{\cal L}_{\phi'}=-\lambda\,\phi' {\bar \chi}\chi=-\lambda\, \phi' ({\bar\chi}_1\chi_1+{\bar\chi}_2\chi_2), 
\eea
In this case, the new singlet scalar does not influence our discussion on the LZ event interpretation, but it provides new annihilation channels for dark matter, $\chi_1{\bar\chi}_1\to \phi'\phi' ,\chi_2{\bar\chi}_2\to \phi'\phi' $. The corresponding cross sections are given by
\bea
\langle\sigma v\rangle_{\chi_1{\bar\chi}_1\to \phi'\phi'} &=&\frac{\lambda^4m^2_\chi (2(m^2_{\phi'}-2m^2_\chi)^2+m^4_\chi)}{24\pi(m^2_{\phi'}-2m^2_\chi)^4}\,\sqrt{1-\frac{m^2_{\phi'}}{m^2_\chi}}\,\cdot\bigg(\frac{6T}{m_\chi}\bigg) \nonumber \\
&\simeq& \langle\sigma v\rangle_{\chi_2{\bar\chi}_2\to \phi'\phi'}.
\eea
Then,  the above additional annihilation channels are suppressed by $p$-wave, so we find that there is no strong constraint on them from indirect detection experiments, even if $\phi'$ decays into the SM particles through its mixing with the SM Higgs.

The relic density for dark matter can be obtained from
\bea
\Omega_\chi h^2=\frac{2.09\times 10^8\,{\rm GeV}^{-1}}{M_P \sqrt{g_*}/x_f(\sigma_s+3\sigma_p/x_f)},
\eea
with $g_*$ being the effective number of relativistic species during freeze-out, $x_f=m_\chi/T_f\simeq 20$ at the freeze-out temperature $T_f$, and $\sigma_s,\sigma_p$ are the $s$-wave and $p$-wave coefficients in the velocity expansion of the total annihilation cross section in the form, $\langle\sigma v \rangle_{\rm tot}=\sigma_0+6\sigma_p/x$, respectively.

In the benchmark scenario with $m_\chi=800\,{\rm GeV}$ and $\Delta m=342\,{\rm keV}$, together with  $m_{Z'}=21(10.5)\,{\rm GeV}$ and $g_{Z'}=0.2(0.1)$, for which  $\sigma_N=6.4\times 10^{-39}\,{\rm cm}^2$, we find that the correct relic density can be obtained for $m_{\phi'}=100\,{\rm GeV}$ and $\lambda=0.71$.

Similarly, in the relatively light DM scenario with $m_\chi=487\,{\rm GeV}$ and $\Delta m=322\,{\rm keV}$, together with $m_{Z'}=16.6(8.3)\,{\rm GeV}$ and $g_{Z'}=0.2(0.1)$, for which $\sigma_N=1.4\times 10^{-38}\,{\rm cm}^2$, we find the consistent parameter space for the relic density for $m_{\phi'}=100\,{\rm GeV}$ and $\lambda=0.55$.

\subsection{Dijet bounds and other collider constraints}

In the $U(1)_{B_3-B_1}$ model, the $Z'$ gauge boson is coupled to the first and third generation quarks at the same time, so dijets can be produced from the Drell-Yann process with $Z'$ exchanges, in particular, at the LHC. Moreover, for $m_{Z'}>2m_t$, a resonant production of $t{\bar t}$ can be also searched for at the LHC.

On the other hand, in the $U(1)_{B_2-B_1}$ model, the $Z'$ gauge boson couples to the first two generation quarks, so dijets are main signatures of the model.

According to the recent analysis on dijet constraints in Ref.~\cite{dijet1,dijet2}, the existing bounds on light dijets for baryonic $Z'$ indicate $g_{Z'}\lesssim 0.1-0.2$ for $m_{Z'}\lesssim 600\,{\rm GeV}$ and there are similar constraints for $m_{Z'}\gtrsim 600\,{\rm GeV}$. Therefore, the parameter space with a light $Z'$ and $g_{Z'}= 0.1-0.2$ for explaining the LZ nuclear recoil is consistent with the current dijet bounds.

There are other potentially important constraints on the model, coming from meson decays in the presence of flavor-violating $Z'$ interactions for quarks. If the quark flavor mixing is attributed to the CKM matrix directly, the flavor violating couplings are suppressed by the CKM mixing angles as in Ref.~\cite{ZpFV}, but the details depend on the concrete realization of the flavor structure. Thus, we postpone this issue to a future work.

\subsection{Indirect detection}

In the inelastic dark matter scenarios where Higgsinos or similar new stable particles with electroweak interactions and a small mass splitting are responsible the LZ nuclear recoil through the $Z$ boson exchange, it has been pointed out that the neutrino searches from IceCube set a strong lower bound on the mass splitting for such an inelastic dark matter beyond $506\,{\rm keV}$, excluding the possibility of explaining the LZ nuclear recoil by inelastic scattering \cite{Pospelov}. The reason is that  Higgsino-like dark matter is captured in the core of the Sun due to strong DM-nucleon scattering and it annihilates into $W^+W^-$ through the $Z$ boson exchange, producing a lot of prompt neutrinos. But, there is no observation of excess neutrinos coming from the Sun according to IceCube data \cite{icecube}.

In our model for inelastic dark matter, there is no coupling between dark matter and $Z$ boson at tree level, but the effective coupling, $Z'-W^+-W^-$, can be generated at one-loop due to quarks coupled to $Z'$ running in loops. Thus, in our model, there is a loop suppression for the production of prompt neutrinos coming from the DM annihilation in the Sun, so the parameter space explaining the LZ data in our model can be consistent  with the IceCube bound \cite{Icecube2}. Furthermore, the annihilation channels in our model, $\chi_1{\bar\chi}_2\to q{\bar q}$ and $\chi_1{\bar\chi}_1\to Z'Z'$ with each $Z'$ decaying by $Z'\to q{\bar q}$, can be sub-dominant for determining the relic density, so the bounds on the excess neutrinos coming from those channels can be satisfied more easily.

More concretely, in the benchmark scenario with $m_\chi=800\,{\rm GeV}$, $\Delta m=342\,{\rm keV}$, $m_{Z'}=21(10.5)\,{\rm GeV}$ and $g_{Z'}=0.2(0.1)$, we obtain the annihilation cross sections for $\chi_1{\bar\chi}_2\to q{\bar q}$, $\chi_1{\bar\chi}_1\to Z'Z'$ as $\langle \sigma v\rangle_{\chi_1{\bar\chi}_2\to q{\bar q}}=2.2(0.13)\times 10^{-29}\,{\rm cm^3/s}$ and $\langle \sigma v\rangle_{\chi_1{\bar\chi}_1\to Z'Z'}=4.5(0.28)\times 10^{-31}\,{\rm cm^3/s}$, respectively, so they are much smaller than the thermal cross section.

Similarly, relatively light DM scenario with $m_\chi=487\,{\rm GeV}$,  $\Delta m=322\,{\rm keV}$, $m_{Z'}=16.6(8.3)\,{\rm GeV}$ and $g_{Z'}=0.2(0.1)$, we also get the annihilation cross sections for $\chi_1{\bar\chi}_2\to q{\bar q}$, $\chi_1{\bar\chi}_1\to Z'Z'$ as $\langle \sigma v\rangle_{\chi_1{\bar\chi}_2\to q{\bar q}}=5.8(0.36)\times 10^{-29}\,{\rm cm^3/s}$ and $\langle \sigma v\rangle_{\chi_1{\bar\chi}_1\to Z'Z'}=1.2(0.076)\times 10^{-30}\,{\rm cm^3/s}$, respectively. Again the DM annihilation cross sections into $q{\bar q}$ and $Z'Z'$ are much smaller than the thermal cross section.

\section{Conclusions}

Wr presented a novel model for inelastic dark matter from a Dirac singlet fermion which is charged under the local $U(1)'$ baryon flavor symmetries. $U(1)_{B_i-B_j}$ with $i\neq j$ and any linear combinations of them are good candidates for the anomaly-free baryon flavor symmetries without extra charged fermions under the SM gauge group.
The baryonic current associated with $U(1)_{B_3-B_1}$ or $U(1)_{B_2-B_1}$ contain at least the first generation quarks for direct detection experiments. A realistic flavor structure for the mixing Yukawa couplings can be realized due to higher dimensional operators or renormalizable couplings between the SM quarks and extra vector-like quarks that are anomaly-free by themselves.  In our model, we obtained a small mass splitting for inelastic dark matter from a small Majorana mass, which is generated only after the $U(1)'$ symmetry is broken spontaneously.

We showed that the LZ excess is accommodated by the DM-nucleon inelastic scattering through a light $Z'$ exchange in our model.
We also found that there is a consistent parameter space for explaining both the LZ nuclear recoil event and the DM relic density and satisfying the existing collider and indirect detection bounds at the same time. In this case, we predicted that it is necessary to take a relatively light $Z'$ boson with $10-20\,{\rm GeV}$ mass and a weak gauge coupling. The future data with higher recoil energies at direct detection experiments will play a decisive role of discriminating between the scenarios of light and heavy dark matter.

We introduced additional DM annihilations for dark matter to get  the correct relic density, being compatible with the LZ event. 
In a simple extension of the model with a new $U(1)'$-neutral light scalar $\phi'$ coupled to dark matter, we showed that the annihilations into the dark sector can also help avoid the potential bounds from direct detection such as neutrinos coming from the annihilations of dark matter captured in the Sun.

\section*{Acknowledgments}

The author thanks Seth Koren on the discussion on the $U(1)'$ baryon flavor symmetry and appreciates support during the visit in CERN by the CERN Scientific Associate.
The work is supported in part by Basic Science Research Program through the National
Research Foundation of Korea (NRF) funded by the Ministry of Education, Science and
Technology (NRF-2022R1A2C2003567).




\end{document}